\documentclass[a4paper,11pt]{article}
\usepackage{graphicx,epsfig,picins}
\usepackage{fancyhdr,fancybox}
\usepackage{indentfirst}
\usepackage{titlesec}
\usepackage{verbatim}
\usepackage[sort&compress, numbers]{natbib}
\usepackage{array,dcolumn,tabularx}
\usepackage{setspace}
\usepackage{geometry}
\usepackage{extarrows,chemarrow,xypic} 
\usepackage[small]{caption2}
\usepackage{lineno}
\usepackage{microtype}
\DisableLigatures[f]{encoding = *, family = * }

\usepackage{times}     

\usepackage{latexsym}
\usepackage{amsmath}                 
\usepackage{amssymb}                 
\usepackage{amsbsy}
\usepackage{amsthm}
\usepackage{amsfonts}
\usepackage{mathrsfs}                
\usepackage{bm}                      
\usepackage{arydshln}
\usepackage{relsize}                 
\usepackage{hyperref}  

\newcommand{\paperfont}{\fontsize{12pt}{1.3\baselineskip}\selectfont}
\begin{document}

\theoremstyle{definition}
\makeatletter
\thm@headfont{\bf}
\makeatother
\newtheorem{definition}{Definition}
\newtheorem{example}{Example}
\newtheorem{theorem}{Theorem}
\newtheorem{lemma}{Lemma}
\newtheorem{corollary}{Corollary}
\newtheorem{remark}{Remark}
\newtheorem{proposition}{Proposition}

\lhead{}
\rhead{}
\lfoot{}
\rfoot{}

\renewcommand{\refname}{References}
\renewcommand{\figurename}{Fig.}
\renewcommand{\tablename}{Table}
\renewcommand{\proofname}{Proof}

\newcommand{\diag}{\mathrm{diag}}
\newcommand{\tr}{\mathrm{tr}}
\newcommand{\dnum}{\mathrm{d}}
\newcommand{\Enum}{\mathbb{E}}
\newcommand{\Pnum}{\mathbb{P}}
\newcommand{\Rnum}{\mathbb{R}}
\newcommand{\Cnum}{\mathbb{C}}
\newcommand{\Znum}{\mathbb{Z}}
\newcommand{\Nnum}{\mathbb{N}}
\newcommand{\F}{\mathscr{F}}
\newcommand{\abs}[1]{\left\vert#1\right\vert}
\newcommand{\set}[1]{\left\{#1\right\}}
\newcommand{\norm}[1]{\left\Vert#1\right\Vert}
\newcommand{\innp}[1]{\langle {#1}]}
\newcommand{\lint}[1]{\left\lfloor#1\right\rfloor}
\newcommand{\rint}[1]{\left\lceil#1\right\rceil}


\title{\textbf{Emergent L\'{e}vy behavior in single-cell stochastic gene expression}}
\author{Chen Jia$^{1}$,\;\;\;Michael Q. Zhang$^{2,3}$,\;\;\;Hong Qian$^4$ \\
\footnotesize $^1$Department of Mathematical Sciences, The University of Texas at Dallas, Richardson, TX 75080, U.S.A.\\
\footnotesize $^2$Department of Biological Sciences, University of Texas at Dallas, Richardson, TX 75080, U.S.A.\\
\footnotesize $^3$MOE Key Lab and Division of Bioinformatics, CSSB, TNLIST, Tsinghua University, Beijing 100084, China\\
\footnotesize $^4$Department of Applied Mathematics, University of Washington, Seattle, WA 98195, U.S.A.}
\date{}                              
\maketitle                           
\thispagestyle{empty}                

\paperfont

\begin{abstract}
Single-cell gene expression is inherently stochastic; its emergent behavior can be defined in terms of the chemical master equation describing the evolution of the mRNA and protein copy numbers as the
latter tends to infinity. We establish two types of ``macroscopic limits": the Kurtz limit is consistent with the classical chemical kinetics, while the L\'{e}vy limit provides a theoretical foundation for an empirical equation proposed in [Phys. Rev. Lett. 97:168302, 2006]. Furthermore, we clarify the biochemical implications and ranges of applicability for various macroscopic limits and calculate a comprehensive analytic expression for the protein concentration distribution in autoregulatory gene networks. The relationship between our work and modern population genetics is discussed.
\\

\noindent 
\end{abstract}

\section*{Introduction}
The mesoscopic stochastic theory of chemical reaction kinetics is a powerful analytic paradigm for single-cell biochemical dynamics \cite{qian2012cooperativity}. At the center of this theory is a limit theorem, first proved by Kurtz in the 1970s \cite{kurtz1972relationship}, which states that when the size $V$ of the reaction vessel tends to infinity, the kinetics of a well-mixed reaction system can be described by a set of ordinary differential equations (ODEs), as intuitively expected from the macroscopic chemical reaction kinetics. It is the macroscopic limit, instead of the mean value, that should be identified as the \emph{emergent behavior} of the stochastic dynamics, as incisively pointed out by Anderson \cite{anderson1972more}: ``It is only as it is considered to be a many body system | in what is often called the $N\rightarrow\infty$ limit | that such [emergent] behavior is rigorously definable." Investigating the limit of $V\to\infty$ or $N\to\infty$, therefore, provides a way to reveal the inherent fundamental character of a stochastic biochemical system.

In general, the stochastic biochemical reaction kinetics has two complementary representations: the stochastic trajectory and the probability distribution. The former is governed by a continuous-time Markov chain that can be simulated via Gillespie's algorithm and the latter is governed by the chemical master equation (CME) first appearing in the work of Delbr\"{u}ck \cite{delbruck1940statistical}. To emphasize this dual perspective, the underlying stochastic dynamics is usually termed the {\em Delbr\"{u}ck-Gillespie process} (DGP) \cite{qian2011nonlinear}.

In recent years, significant progress has been made in the kinetic theory of single-cell stochastic gene expression based on the central dogma of molecular biology \cite{peccoud1995markovian, paulsson2000random, kepler2001stochasticity, paulsson2005models, friedman2006linking, shahrezaei2008analytical, kumar2014exact, newby2015bistable, ge2015stochastic, lin2016gene, bressloff2017stochastic, jia2017stochastic}. A thorough study based on the DGP framework, in terms of the protein copy number, was carried out by Shahrezaei and Swain \cite{shahrezaei2008analytical}. However, in bulk experiments and many single-cell experiments without single-molecule resolution such as RNA sequencing and flow cytometry, data are usually obtained as continuous variables at a macroscopic scale. At the center of the kinetic theory in terms of the \emph{protein concentration} is an empirical equation proposed by Friedman, Cai, and Xie (FCX) \cite{friedman2006linking}. However, the mathematical foundation of the now classical FCX equation still remains unclear. This paper addresses its theoretical foundation.

\section*{Emergent behavior in single-cell stochastic gene expression}
\begin{figure}[!thb]
\begin{center}
\centerline{\includegraphics[width=0.6\textwidth]{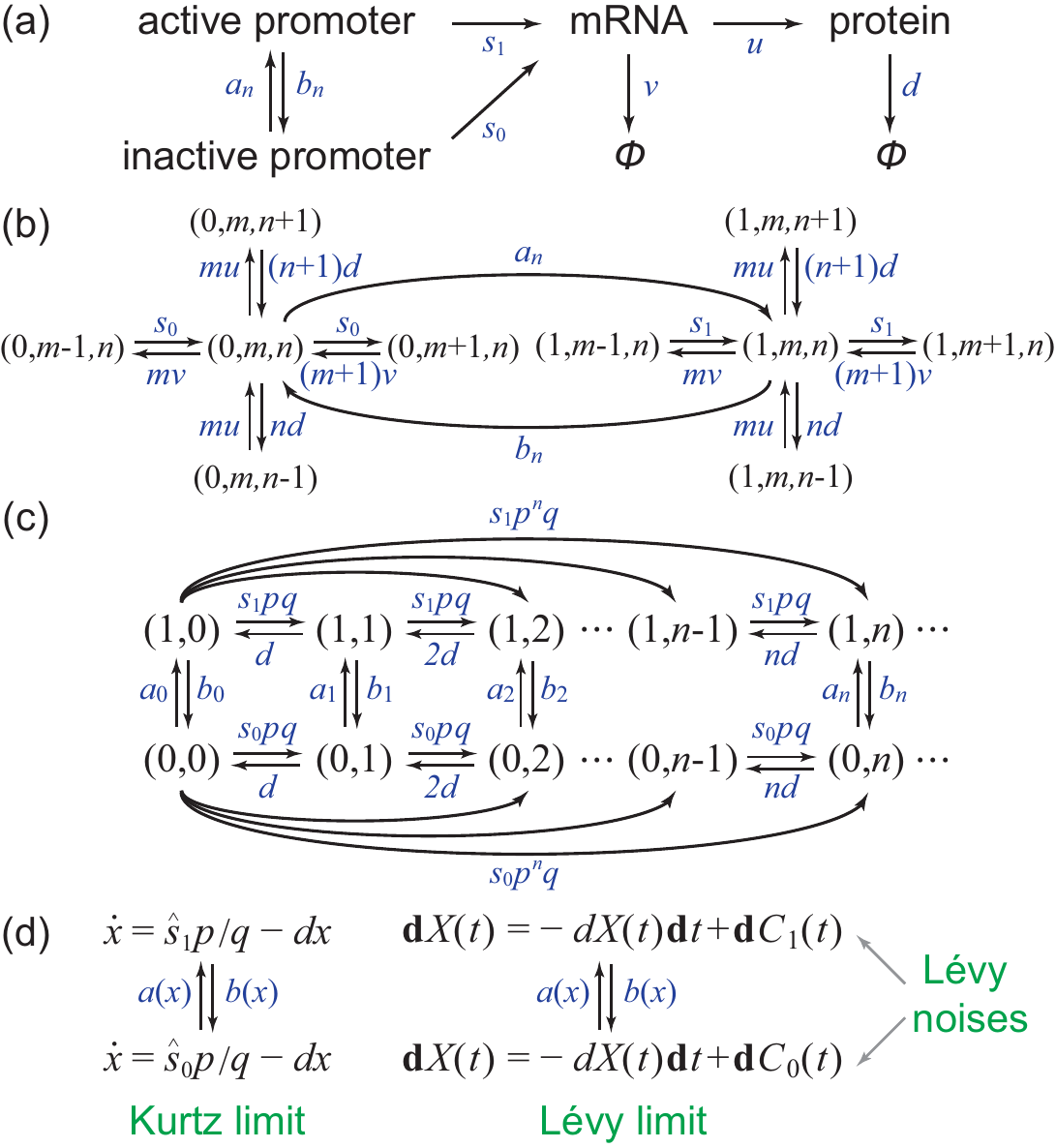}}
\caption{(a) The canonical three-stage representation of stochastic gene expression.
(b) The transition diagram of the DGP.
(c) The transition diagram of the reduced model when $\epsilon\ll 1$.
(d) The Kurtz and L\'{e}vy limits.}\label{threestage}
\end{center}
\end{figure}

We consider the canonical three-stage representation of stochastic gene expression in a single cell with size $V$, with $V\to\infty$ corresponding to a macroscopic scale, as illustrated in Fig. \ref{threestage}(a) \cite{shahrezaei2008analytical}. The size $V$ in chemistry stands for the reaction volume \cite{kurtz1972relationship}, but in molecular biology it could also be the maximum protein copy number \cite{kepler2001stochasticity}, etc. The biochemical state of the gene of interest can be described by three variables: the promoter activity $i$ with $i = 1$ and $i = 0$ corresponding to the active and inactive states of the promoter, respectively, the mRNA copy number $m$, and the protein copy number $n$. Then the kinetics can be described by the DGP depicted in Fig. \ref{threestage}(b). Here $s_1$ and $s_0$ are the transcription rates when the promoter is active and inactive, respectively; $u$, $v$, and $d$ are the rate constants for translation, mRNA degradation, and protein degradation, respectively; $a_n$ and $b_n$ are the switching rates of the promoter between the active and inactive states \cite{chong2014mechanism}. In living cells, the products of many genes also regulate their own expression to form an autoregulatory gene network. This suggests that the promoter switching rates $a_n$ and $b_n$ generally depend on the protein copy number $n$.

Experimentally, it has been consistently observed that the mRNA decays substantially faster than its protein counterpart \cite{shahrezaei2008analytical}. Then the process of protein synthesis followed by mRNA degradation is essentially instantaneous: Protein synthesis in single cells occurs in random bursts \cite{cai2006stochastic}. Once an mRNA is synthesized, it can either produce a protein with probability $p = u/(u+v)$ or be degraded with probability $q = v/(u+v)$. Thus the probability that $j$ proteins are synthesized in a single burst will be $p^jq$, which follows the geometric distribution \cite{berg1978model}. The average number of proteins synthesized per mRNA, also called the mean burst size, is then $\sum_{j=0}^\infty jp^jq = p/q$. These considerations yield the reduced Markov model illustrated in Fig. \ref{threestage}(c) \cite{paulsson2000random}.

In fact, the reduced model can be derived rigorously from the original DGP. To do this, let $\epsilon = d/v$ denote the ratio of the mRNA and protein lifetimes. Let $q_{(i,m,n)}$ denote the rate at which the system leaves state $(i,m,n)$, which is defined as the sum of transition rates from state $(i,m,n)$ to other states \cite{supp}. Since $\epsilon\ll 1$, we say that $(i,m,n)$ is a fast state if $q_{(i,m,n)}\rightarrow\infty$ as $\epsilon\rightarrow 0$. Otherwise, $(i,m,n)$ is called a slow state. If $(i,m,n)$ is a fast state, then the time that the system stays in this state will be very short. By a recently developed simplification method of two-time-scale Markov chains \cite{jia2016reduction, jia2016simplification, jia2017simplification}, the DGP can be simplified by removal of all the fast states. It is easy to check that
\begin{equation*}
\begin{split}
q_{(0,m,n)} &= md(u/v+1)\epsilon^{-1}+a_n+s_0+nd,\\
q_{(1,m,n)} &= md(u/v+1)\epsilon^{-1}+b_n+s_1+nd.
\end{split}
\end{equation*}
This indicates that all the states $(i,m,n)$ with $m\geq 1$ are fast states and can be removed and only the states $(i,0,n)$ are retained. Thus the original DGP can be simplified to the reduced model with effective transition rates depicted in Fig. \ref{threestage}(c) \cite{supp}. In the reduced model, the biochemical state of the gene is only described by the variables $i$ and $n$. It yields large increments of the protein number, which suggests that protein synthesis occurs in random bursts.

Let $\alpha(t)$ and $N(t)$ denote the promoter activity and protein copy number in a single cell at time $t$, respectively. Then $X_V(t) = N(t)/V$ stands for the protein concentration. When $\epsilon\ll 1$, $(\alpha(t),N(t))$ can be described by the reduced model. Thus $(\alpha(t),X_V(t))$ is a Markov chain with state space $\{(i,n/V): i = 0,1,n=0,1,2\cdots\}$. Under mild conditions, the evolution of a Markov process is uniquely determined by its generator. In particular, the generator $\mathcal{A}_V$ of the Markov chain $(\alpha(t),X_V(t))$ is given by
\begin{equation*}\left\{
\begin{split}
& \mathcal{A}_Vf_1\left(\tfrac{n}{V}\right) =
nd\Big[f_1\left(\tfrac{n}{V}-\tfrac{1}{V}\right)-f_1\left(\tfrac{n}{V}\right)\Big]
+b_n\Big[f_0\left(\tfrac{n}{V}\right)\\
&\; -f_1\left(\tfrac{n}{V}\right)\Big]+\sum_{j=0}^\infty s_1p^jq\Big[f_1\left(\tfrac{n}{V}+\tfrac{j}{V}\right)-f_1\left(\tfrac{n}{V}\right)\Big],\\
& \mathcal{A}_Vf_0\left(\tfrac{n}{V}\right) = nd\Big[f_0\left(\tfrac{n}{V}-\tfrac{1}{V}\right)-f_0\left(\tfrac{n}{V}\right)\Big]
+a_n\Big[f_1\left(\tfrac{n}{V}\right)\\
&\; -f_0\left(\tfrac{n}{V}\right)\Big] +\sum_{j=0}^\infty s_0p^jq\Big[f_0\left(\tfrac{n}{V}+\tfrac{j}{V}\right)-f_0\left(\tfrac{n}{V}\right)\Big].
\end{split}\right.
\end{equation*}

Let $x = n/V$ and $y = j/V$ and let $a(x) = a_n$ and $b(x) = b_n$. Under the framework of mesoscopic chemical reaction kinetics, $\textrm{DNA}\rightarrow\textrm{mRNA}$ is a zero-order reaction and thus the transcription rate should scale with size $V$, that is, $s_i = \hat{s}_iV$ \cite{kurtz1972relationship}. As $V\rightarrow\infty$, the generator $\mathcal{A}_V$ will converge to another operator $\mathcal{B}$:
\begin{equation*}\left\{
\begin{split}
\mathcal{B}f_1(x) =&\; (\hat{s}_1p/q-dx)f_1'(x)+b(x)\big[f_0(x)-f_1(x)\big],\\
\mathcal{B}f_0(x) =&\; (\hat{s}_0p/q-dx)f_0'(x)+a(x)\big[f_1(x)-f_0(x)\big].
\end{split}\right.
\end{equation*}
This shows that the discrete-valued Markov chain $(\alpha(t),X_V(t))$ will converge to a continuous-valued Markov process $(\alpha(t),X(t))$ with generator $\mathcal{B}$. Mathematically, the limiting process is a piecewise deterministic Markov process (PDMP), as illustrated in Fig. \ref{threestage}(d). This macroscopic limit will be named as the Kurtz limit because it is consistent with the classical chemical kinetics: Given a particular promoter state, the protein concentration evolves as an ODE with no fluctuations. The PDMP was introduced in \cite{newby2015bistable, ge2015stochastic} for studying stochastic phenotype switching. In \cite{lin2016gene}, Lin and Doering considered a gene network with positive autoregulation and obtained the PDMP by taking a different but mathematically equivalent limit. Recently, there has been many studies on gene expression kinetics based on the PDMP model and the detailed analysis can be found in \cite{newby2015bistable, bressloff2017stochastic}.

Interestingly, there is another macroscopic limit that is more consistent with single-cell experiments.
To see this, we assume that the mean burst size $p/q = V/k$ scales with size $V$. Here we shall treat $s_i$ and $k$ as constants and take the limit $V\rightarrow\infty$. Under these assumptions, we have $p\rightarrow 1$, $qV\rightarrow k$, and $p^j = e^{j\log(1-q)}\rightarrow e^{-ky}$. Thus the generator $\mathcal{A}_V$ will converge to a different operator $\mathcal{A}$:
\begin{equation*}\left\{
\begin{split}
\mathcal{A}f_1(x) =&\; -dxf_1'(x)+b(x)\big[f_0(x)-f_1(x)\big]\\
&+s_1\int_0^\infty ke^{-ky}\big[f_1(x+y)-f_1(x)\big]\dnum y,\\
\mathcal{A}f_0(x) =&\; -dxf_0'(x)+a(x)\big[f_1(x)-f_0(x)\big]\\
&+s_0\int_0^\infty ke^{-ky}\big[f_0(x+y)-f_0(x)\big]\dnum y.
\end{split}\right.
\end{equation*}
This shows that the Markov chain $(\alpha(t),X_V(t))$ will converge to a different Markov process $(\alpha(t),X(t))$ with generator $\mathcal{A}$. Mathematically, the limiting process is a switching (hybrid) stochastic differential equation (SDE) driven by L\'{e}vy noises, as illustrated in Fig. \ref{threestage}(d). Here $C_i(t)$ is a compound Poisson process, a particular kind of L\'{e}vy process, with arrival rate $s_i$ and jump distribution $w(x) = ke^{-kx}$. This can be explained as follows. When the promoter is in state $i$, the process of mRNA synthesis can be described by a Poisson process with arrival rate $s_i$ and each mRNA can produce proteins with the burst size having the exponential distribution $w(x)$, which can be viewed as the continuous limit of the geometric distribution. Thus the process of protein synthesis can be described by the compound Poisson process $C_i(t)$. We shall name this macroscopic limit as the L\'{e}vy limit. Given a particular promoter state, the protein concentration still evolves as a stochastic process with large fluctuations.

\begin{figure}[!htb]
\begin{center}\includegraphics[width=0.7\textwidth]{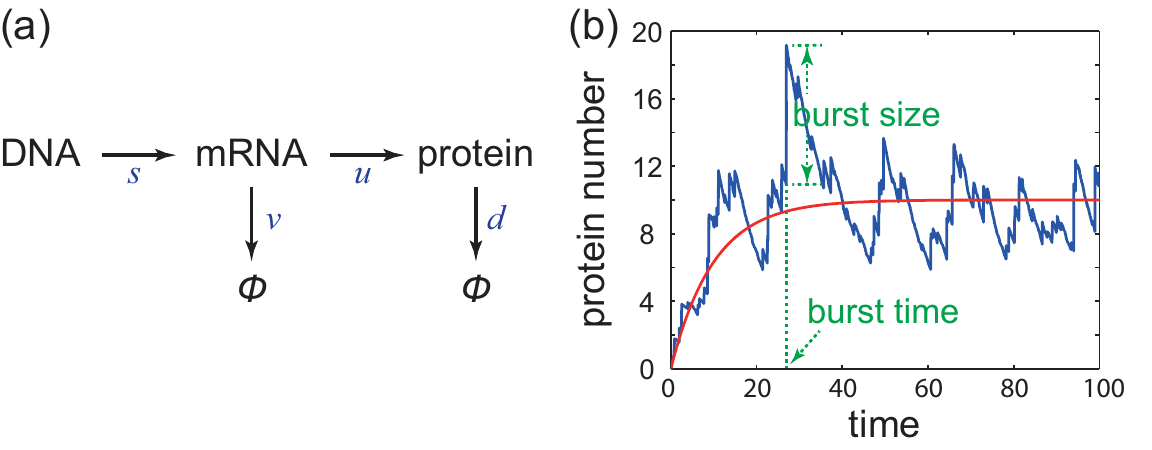}
\caption{(a) The two-stage representation of stochastic gene expression.
(b) Trajectories of the Kurtz (red) and L\'{e}vy (blue) limits. In the simulation, the model parameters are chosen as $s = 1$, $d = 0.1$, $k = 1$, and $V = 50$.}\label{twostage}
\end{center}
\end{figure}

Fig. \ref{twostage}(b) illustrates the simulated trajectories of the two kinds of macroscopic limits when the promoter is always active, that is, $b_n = 0$. It can be seen that the trajectories of the Kurt limit are continuous, while the L\'{e}vy limit has discontinuous trajectories. The jump point of the trajectory corresponds to the burst time and the jump height corresponds to the burst size. For any fixed protein concentration $x$, consider the transition rate $q_{(i,n),(i,n+xV)} = s_ip^{xV}q$ of the reduced model from state $(i,n)$ to $(i,n+xV)$, where $xV$ is assumed to be an integer for simplicity. Under the assumption of the Kurtz limit, $q_{(i,n),(i,n+xV)} = \hat{s}_iqVp^{xV}$, which decays to zero at exponential speed. Under the assumption of the L\'{e}vy limit, $q_{(i,n),(i,n+xV)} \approx s_ike^{-kx}/V$, which decays to zero with a power law. Thus the L\'{e}vy limit allows a larger probability to yield large increments. This explains why the trajectories of the L\'{e}vy limit are discontinuous.

Let $p_i(x)$ denote the probability density of the protein concentration when the promoter is in state $i$ and let $p(x) = p_0(x)+p_1(x)$ denote the total probability density of the protein concentration. Then the evolution of the L\'{e}vy limit is governed by the Kolmogorov forward equation $\partial_tp_i(x) = \mathcal{A}^*p_i(x)$, that is,
\begin{equation}\label{fokkerplanck}\left\{
\begin{split}
\partial_tp_1(x) =&\; d\partial_x\big(xp_1(x)\big)+s_1\int_0^xke^{-k(x-y)}p_1(y)\dnum y\\
&+a(x)p_0(x)-\big[b(x)+s_1\big]p_1(x), \\
\partial_tp_0(x) =&\; d\partial_x\big(xp_0(x)\big)+s_0\int_0^xke^{-k(x-y)}p_0(y)\dnum y\\
&+b(x)p_1(x)-\big[a(x)+s_0\big]p_0(x),
\end{split}\right.
\end{equation}
where $\mathcal{A}^*$ is the adjoint of $\mathcal{A}$. Based on this equation, we can obtain a general form of the steady-state distribution of the protein concentration, instead of the protein copy number as discussed in \cite{shahrezaei2008analytical}, in autoregulatory networks. For simplicity, we assume that the promoter switching rates have the form of $a_n = a$ and $b_n = b+\gamma n$ \cite{kumar2014exact}, where $b$ is the spontaneous switching rate from the active to the inactive states and $\gamma$ is the feedback strength. This model can be used to analyze networks with either positive or negative autoregulation. When $s_1 > s_0$, the feedback term $\gamma n$ inhibits protein synthesis and leads to negative feedback. In contrast, $s_1 < s_0$ leads to positive feedback. In autoregulatory networks, the steady-state protein distribution is given by $p_{ss}(x) = u\ast v(x)$ \cite{supp}, where $\ast$ denotes the convolution,
\begin{equation*}
u(x) = \frac{k^{s_0/d}}{\Gamma(s_0/d)}x^{s_0/d-1}e^{-kx},\\
\end{equation*}
is the Gamma distribution and
\begin{equation*}
\begin{split}
v(x) = \;&\frac{\Gamma(\beta)}{\Gamma(\alpha_1)\Gamma(\alpha_2){}_2F_1(\alpha_1,\alpha_2;\beta;\gamma/dw)}\\
& we^{(\frac{\gamma}{d}-\frac{w}{2})x}(wx)^{\frac{\alpha_1+\alpha_2-3}{2}}
W_{\frac{\alpha_1+\alpha_2+1}{2}-\beta,\frac{\alpha_1-\alpha_2}{2}}(wx).
\end{split}
\end{equation*}
Here ${}_2F_1(\alpha_1,\alpha_2;\beta;x)$ is the Gaussian hypergeometric function, $W_{\alpha,\beta}(x)$ is the Whittaker function, and
\begin{equation*}
\begin{split}
&\alpha_1+\alpha_2 = \frac{a+b+s_1-s_0}{d},\;\;\;\alpha_1\alpha_2 = \frac{a(s_1-s_0)}{d^2},\\
&\beta = \frac{\gamma s_1+(a+b)(dk+\gamma)}{d(dk+\gamma)},\;\;\;w = k+\frac{\gamma}{d}.
\end{split}
\end{equation*}
When $a = 0$ or $s_1 = s_0$, there is only one promoter state and we have $\alpha_1 = 0$ and $v(x) = \delta(x)$. In this case, the protein concentration has the Gamma distribution: $p_{ss}(x) = u\ast\delta(x) = u(x)$ \cite{friedman2006linking}. If we assume that the mRNA produces no proteins when the promoter is inactive, that is, $s_0 = 0$, we have $u(x) = \delta(x)$ and thus the protein concentration has the Whittaker-type distribution: $p_{ss}(x) = \delta\ast v(x) = v(x)$.

The emergent L\'{e}vy behavior of single-cell stochastic gene expression is itself a stochastic process with large fluctuations, which shows that the stochastic effects cannot be averaged out at the macroscopic scale. This provides a mechanistic foundation, from the viewpoint of many-body theoretical physics, for intracellular variations at the epigenetic and phenotypic level. The L\'{e}vy limit of the DGP is on par with the Feller-Kimura diffusion limit of the Wright-Fisher random mating model \cite{ewens-book}, which has become the theoretical foundation for ``nearly all of modern population genetics" \cite{wakeley2005limits}. As a comparison, the DGP and the Wright-Fisher model are both discrete-valued Markov chains, while the L\'{e}vy limit of the former and the diffusion limit of the latter are both continuous-valued Markov processes. However, they are subtly different because the diffusion limit has continuous trajectories, while the L\'{e}vy limit has discontinuous ones: The intracellular diversity is much greater. This insight may have far-reaching implications to many biological phenomena such as bacterial drug resistance and non-genetic cancer heterogeneity \cite{huangsui}. The full comparison between our theory and the theory of population genetics is listed in Table \ref{comparison}.
\begin{table}[!htb]\centering
\begin{tabular}{|c|c|c|} \hline
                 & Epigenetics      & Population genetics \\ \hline
Discrete model   & DGP              & Wright-Fisher model \\ \hline
Continuous model & L\'{e}vy limit   & Feller-Kiruma limit \\ \hline
Noise term       & L\'{e}vy process & Brownian motion \\ \hline
Trajectories     & Discontinuous    & Continuous \\ \hline
\end{tabular}
\caption{Comparison between our theory and the theory of population genetics.}\label{comparison}
\end{table}

\section*{Two special cases}
There are two special scenarios that is most interesting. The first one occurs when the promoter is always active, that is, $b_n = 0$. In this case, stochastic gene expression in a single cell has the two-stage representation illustrated in Fig. \ref{twostage}(a), where $s$ is the transcription rate. If the transcription rate $s = \hat{s}V$ scales with size $V$, the two-stage model has the following Kurtz limit, which is an ODE:
\begin{equation}\label{ODE}
\dot{x} = -dx+\hat{s}p/q.
\end{equation}
where $\hat{s}p/q$ is the mean synthesis rate of the protein. If the mean burst size $p/q = V/k$ scales with size $V$, however, the two-stage model has the following L\'{e}vy limit, which is an SDE driven by L\'{e}vy noise:
\begin{equation*}
\dnum X(t) = -dX(t)\dnum t + \dnum C(t),
\end{equation*}
where $C(t)$ is a compound Poisson process with arrival rate $s$ and jump distribution $w(x) = ke^{-kx}$.
From Eq. \eqref{fokkerplanck}, the evolution of the L\'{e}vy limit is governed by
\begin{equation*}
\partial_tp(x) = d\partial_x\big(xp(x)\big)+s\int_0^xw(x-y)p(y)\dnum y-sp(x).
\end{equation*}
This is exactly the empirical equation proposed by FCX \cite{friedman2006linking}, in which the authors made clear that $w(x-y)$ stands for the transition probability of the protein concentration from $y$ to $x$ in a single burst. They further combined experimental observations \cite{cai2006stochastic, yu2006probing} to show that the burst size $x-y$ has an exponential distribution $w(x-y)$. Our theory shows that the classical FCX equation can be derived theoretically from the fundamental single-cell biochemical reaction kinetics without resorting to experimental information.

To further compare the two kinds of limits, we introduce the Laplace transform $f(\lambda) = \int_0^\infty p(x)e^{-\lambda x}\dnum x$. Then the FCX equation is converted to the first-order linear partial differential equation
\begin{equation*}
\partial_t f = -d\lambda\partial_\lambda f-\frac{s\lambda f}{\lambda+k}.
\end{equation*}
It is easy to see that the mean protein concentration $\langle x\rangle$ can be recovered from $f(\lambda)$ as $\langle x\rangle = -\partial_\lambda f(0)$. Thus the evolution of $\langle x\rangle$ is governed by the following ODE:
\begin{equation}\label{mean}
\frac{\dnum\langle x\rangle}{\dnum t} = -d\langle x\rangle+\frac{s}{k}.
\end{equation}
Since $s = \hat{s}V$ and $p/q = V/k$, we have $\hat{s}p/q = s/k$. Comparing Eqs. \eqref{ODE} and \eqref{mean}, we clearly see that the Kurtz limit is exactly the mean of the L\'{e}vy limit, as illustrated in Fig. \ref{twostage}(b).

The biochemical implications of the two kinds of limits can be seen as follows. Recall that the mean protein copy number in a single cell is the product of the mean burst frequency $s/d$ and the mean burst size $p/q$. If $s/d\gg p/q$, the Kurtz limit is valid. This condition is consistent with bulk experiments in which a large number of cells are ground to form a cell extract and thus the DNA copy number is very large. If $p/q\gg s/d$, the L\'{e}vy limit is applicable. In living cells, the mean burst size $p/q$ is relatively large, typically on the order of 100 for an \emph{E. coli} gene \cite{paulsson2005models}. Thus this condition corresponds to single-cell experiments in which the DNA copy number is very small.

We stress here that our theory can be also applied to model stochastic mRNA expression with transcriptional bursts. Recent bulk \cite{marguerat2012coordinating} and single-cell \cite{padovan2015single} experiments have shown that mRNA abundances in individual eukaryotic cells generally scale with cellular volume. Single-molecule imaging techniques \cite{padovan2015single} have further shown that cellular volume affects mRNA abundances through modulation of transcriptional burst size. As a result, the L\'{e}vy limit is also applicable to describe mRNA fluctuations in single cells with large volumes.

Many previous studies also focused on the scenario when the promoter switches rapidly between the active and inactive states, that is, $a_n,b_n\gg s_1,d$ \cite{friedman2006linking}. Under this assumption, the protein concentration will reach a quasi-steady state between the active and inactive states, which suggests that
\begin{equation*}
\begin{split}
p_1(x)\approx a(x)p(x)/(a(x)+b(x)),\\
p_0(x)\approx b(x)p(x)/(a(x)+b(x)).
\end{split}
\end{equation*}
From Eq. \eqref{fokkerplanck}, the evolution of the L\'{e}vy limit is governed by
\begin{equation*}
\begin{split}
\partial_tp(x) =&\; d\partial_x\big(xp(x)\big)+\int_0^xke^{-k(x-y)}c(y)p(y)\dnum y-c(x)p(x),
\end{split}
\end{equation*}
where $c(x) = (a(x)s_1+b(x)s_0)/(a(x)+b(x))$ is the effective transcription rate. This empirical equation has also appeared in \cite{friedman2006linking} and here we provide a theoretical foundation of this equation as the emergent behavior of the fundamental biochemical reaction kinetics. In this case, the L\'{e}vy limit is no longer an SDE driven by L\'{e}vy noise. However, it falls into the category of L\'{e}vy-type processes \cite{applebaum2009Levy}, which behave locally like L\'{e}vy processes. As a summary, we list all kinds of macroscopic limits and their ranges of applicability in Table \ref{requirements}.
\begin{table}[!htb]\centering
\begin{tabular}{|c|c|} \hline
Macroscopic limits             & Ranges of applicability            \\ \hline
ODE                            & $s_1/d\gg p/q$, $b_n = 0$          \\ \hline
PDMP                           & $s_1/d\gg p/q$                     \\ \hline
L\'{e}vy-driven SDE            & $p/q\gg s_1/d$, $b_n = 0$          \\ \hline
Switching L\'{e}vy-driven SDE  & $p/q\gg s_1/d$                     \\ \hline
L\'{e}vy-type process          & $p/q\gg s_1/d$, $a_n,b_n\gg s_1,d$ \\ \hline
\end{tabular}
\caption{Macroscopic limits of single-cell gene expression kinetics and their ranges of applicability.}\label{requirements}
\end{table}

\section*{Conclusions}
We show that deterministic Kurtz and stochastic L\'{e}vy behaviors naturally emerge from the fundamental single-cell gene expression kinetics. When the transcription rate scales with size, the macroscopic limit is a PDMP which is consistent with the classical deterministic chemical kinetics in aqueous solution. When the mean burst size scales with size, however, the macroscopic limit is a switching L\'{e}vy-driven SDE which captures intracellular variations at the epigenetic level. The L\'{e}vy limit provides a theoretical foundation for the classical FCX empirical equation and gives by far the most general form for the steady-state distribution of the protein concentration. Our theory unifies various continuous gene expression models proposed in the previous literature and clarifies their biochemical implications and ranges of applicability.

\section*{Acknowledgements}
We are grateful to the anonymous referees for their valuable comments and suggestions which helped us greatly in improving the quality of this paper. This work was supported by NIH Grants MH102616, MH109665, and R01GM109964, and also by NSFC Grants 31671384 and 91329000.

\setlength{\bibsep}{5pt}
\small\bibliographystyle{nature}
\bibliography{Levy}
\end{document}